# Long-Range Attraction between Graphene and Water/Oil Interfaces

Avishi Abeywickrama,[1] Douglas H. Adamson,[2] Hannes C. Schniepp[1,*]

1) Department of Applied Science, William & Mary, Virginia, USA

2) Institute of Materials Science Polymer Program and Department of Chemistry, University of Connecticut, Storrs, Connecticut, USA

*) schniepp@wm.edu

## Abstract

We directly measured the interactions between a hydrophobic solid and a hydrophobic liquid separated by water using force spectroscopy, where colloidal probes were coated with graphene oxide (GO) to interact with immobilized heptane droplets in water. We detected attractions with a long range of ~0.5 µm, which cannot be readily explained by standard Derjaguin–Landau–Verwey–Overbeek (DLVO) theory. When the GO was reduced to become more hydrophobic, these forces increased in strength and ranged up to 1.2 µm, suggesting that the hydrophobic nature of the involved surfaces critically influences the observed long-range forces. Previous studies have addressed such hydrophobic attractions, but were limited to solid/water/solid and solid/water/air scenarios. Here we expand this knowledge to include the solid/water/liquid situation. Based on our results, we propose air bubbles attached to the colloidal probe and molecular rearrangement at the water/oil interface as possible origins of the observed interactions. The proposed mechanism expands insights gained from previous to the solid/water/liquid situation and is universally applicable to describe attractive interactions between hydrophobic bodies of any kind separated by water. Our work will be useful to understand and motivate the formation of many colloid and interface phenomena, including emulsions using 2D materials and other amphiphilic/hydrophobic particles.

## Introduction

Interactions between mesoscopic hydrophobic surfaces in water with an interaction range beyond what can be explained in standard Derjaguin–Landau–Vervey–Overbeek (DLVO) theory have been controversially discussed in the literature over the last 20 years.[1] Different interaction ranges varying from ~10 nm to >100 nm, depending on the system and the measurement method, have been reported.[2,3] Several mechanisms have been suggested to explain these interactions, including electrostatic effects,[4,5] correlated charge fluctuations,[6] correlated dipole interactions,[7] and bridging of air bubbles attached to the hydrophobic surfaces.[8,9] These air bubbles can be hundreds of nanometers in size, thus leading to long-range attraction between the surfaces.[10] All these long-range interactions have also been termed "extrinsic" hydrophobic forces,[11] because they are different from the "true" or "intrinsic" hydrophobic interactions, featuring much shorter interaction ranges of only a few Ångstroms/nanometers and are due to structuring of water molecules near the hydrophobic entity.[12,13] In this work, we focus exclusively on studying the long-range extrinsic hydrophobic interactions. A better understanding of these relatively complex forces is needed, because they play an essential role in many applications involving colloid and interface phenomena, including air flotation of mineral particles and bitumen droplets,[14,15] interfacial assembly of nanoparticles,[16] preparation of multifunctional foams and nanocomposites,[17] controlled drug delivery, therapeutics,[18] agglomeration of nonpolar particles in water,[1,19] oil/water separation (using Janus membranes)[20–22] and gas transportation.[23]

In the earliest experimental studies of long-range hydrophobic forces, only attractions between two hydrophobic solids in water were considered (Figure 1A).[8,24–26] These experiments were later expanded to replace one solid surface with an air bubble (Figure 1B), as water/air interfaces are generally considered hydrophobic.[27–29] However, there are still no studies of interactions in water between a solid and a hydrophobic *liquid* (Figure 1C). Here, we directly measure these interactions between a hydrophobic solid surface and a water/hydrophobic liquid interface, also referred to as the water/oil interface (Figure 1D). Studying this scenario involving one liquid hydrophobic body, is particularly important to further understand particle stabilized emulsions[30] that are widely applied in food technology,[31] cosmetic products,[32] oil recovery,[33] and drug delivery.[34]

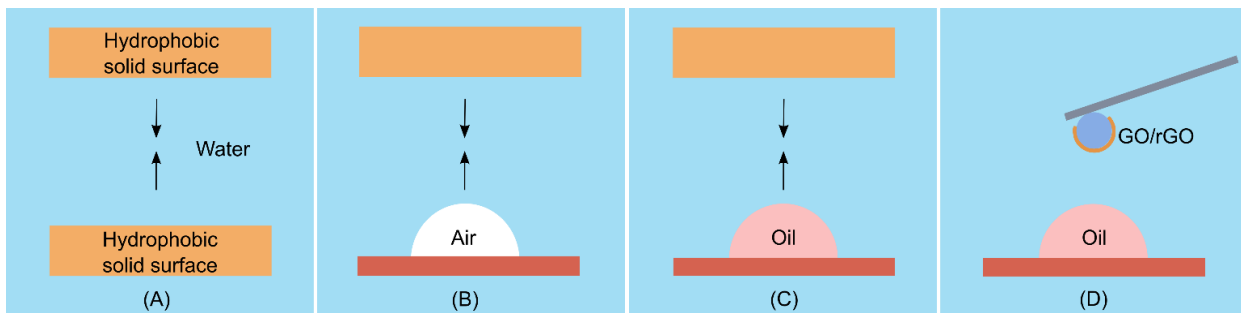

Figure 1. Interactions between hydrophobic bodies in water: (A) two solid surfaces, (B) a solid surface and air, (C) a solid surface and a liquid (oil). (D) Force measurements between graphene oxide (GO) / reduced graphene oxide (rGO) and a water/oil interface.

Different techniques have been used to study interactions at the liquid/liquid interface, such as the surface force apparatus (SFA),[16,35] interferometry, and total internal reflection microscopy (TIRM).[36] However, these techniques have several limitations, i.e., the liquid droplet must be in the millimeter scale and the substrates must be transparent. Force spectroscopy (FS)[37] facilitates measuring interactions between microscale bodies with pN force resolutions and does not require transparent substrates. This study uses FS to directly measure interactions between a solid microparticle mounted to an atomic force microscope (AFM) probe and the water/oil interface (Figure 1D). While FS has been used to investigate interactions at the water/air interface,[29,38,39] only a few studies have been conducted using this technique at the liquid/liquid interface. Force spectroscopy at the water/oil interface has been demonstrated by Dagastine et al.,[40] who studied immobilized tetradecane droplets in water. We advanced this approach by producing a large number of immobilized heptane droplets with well-defined diameters in an aqueous milieu. This allowed us to study interactions between the water/oil interface and a solid surface with varying hydrophobicity levels for the first time. A spherical particle with a 7 μm diameter was first attached to the AFM cantilever and then coated with graphene oxide (GO), as shown in Figure 1D. This approach provided a well-defined surface geometry amenable to modeling, with tunable hydrophobicity levels because the surface energy of GO can be continuously tuned via photoreduction.

The interactions of graphene and related compounds with the water/oil interface are particularly interesting since this interface has been exploited for exfoliation and separation for these layered 2D materials. Our approach thus also enables us to study this technologically important phenomenon systematically. For example, Biswas et al.[41] and Woltornist et al.[42] have trapped graphene sheets at the chloroform/water and heptane/water interfaces, respectively. Similarly, GO has shown an affinity to the water/oil interface.[43–47] Though initially considered hydrophilic, recent work has emphasized the amphiphilic nature of GO,[46] with hydrophilic edges and more hydrophobic basal planes driving assembly at the water/oil interface. Also, Thickett et al.[48] have found that lowering the oil's polarity increases GO sheets' adsorption to the water/oil interface. While the affinity and binding energetics of graphene and GO with the water/oil interface can generally be understood based on interfacial energies,[19] knowing the interaction range of the forces between GO/graphene and the water/oil interface will influence the *kinetics* of the interfacial adsorption and separation process.

We used heptane as the oil phase for our FS measurements, mimicking the system that Woltornist et al.[42] successfully used to synthesize graphene. After acquiring FS measurements, we converted force–displacement curves into force–distance curves to study the interaction ranges. They allowed us to learn more about the nature and underlying mechanisms of observed hydrophobic interactions.

## Methods

**Substrate preparation and droplet immobilization.** Functionalized silicon substrates were used to immobilize the heptane droplets in water. Pieces of a silicon wafer (Ted Pella, Type P) were first cleaned using the following three steps. (1) Sonication in surfactant solution using a bath sonicator (Fisher Scientific FS30 ultrasonic cleaner) for 30 minutes at 60 °C. (2) Rinsing with DI water (Millipore Synergy water purification system), followed by repeating step 1 in DI water without surfactant. (3) Exposing the substrates to high-intensity UV light in a UV ozone cleaner (Novascan PSPD) for 30 minutes, then leaving them for 30 minutes in the ozone environment. After cleaning, the substrates were functionalized with (tridecafluoro-1,1,2,2-tetrahydrooctyl) trichlorosilane (Gelest) — from now on simply referred to as "silane" — to make them hydrophobic, as shown in Figure 2A. Therefore, TEM grids (Veco Hexagonal 200 mesh Copper grid, 3.05 mm diameter, 0.8 mm thickness) were first laid flat on the substrates as masks (Figure 2A(i)), then placed in a vacuum desiccator with a droplet of silane on a separate glass slide for 2 hours (Figure 2A(ii)). This functionalized the substrates with a molecularly thin layer of hydrophobic silane, except for areas covered by the TEM grids, which remained hydrophilic. The TEM grids were then removed from the silicon by immersion in water (Figure 2A(iii)). Care was taken to ensure they did not slide over the newly functionalized regions to avoid scratching or other damage. Ultimately, this procedure created substrates with hydrophobic patches (Figure 2A(iv)), which were used to pin heptane droplets. The pinned droplets were obtained by letting heptane run along the substrates and quickly immersing them in the water-filled AFM liquid cell.

**Fabrication of rGO coated colloidal probes.** Epoxy glue (Ace Hardware, Marine epoxy, 25 minutes set time, 24 hours curing time) was used to attach silica spheres (Bangs Laboratories, 7 µm diameter) onto tipless silicon cantilevers (Budget Sensors, All-in-one Tipless uncoated) featuring a nominal spring constant of 0.2 N/m. The spheres were placed on a glass substrate via spin coating from aqueous suspension. A micromanipulator (Newport Corporation XYZ translation stage, model #: 460-XYZ), attached to an inverted microscope (Olympus IX71), was used for the precise positioning of the cantilever. It was used to apply epoxy at the end of the cantilever and then to move it to one of the silica spheres for pick-up. The epoxy was then allowed 24 hours to cure. The sphere surface topography was characterized using contact-mode AFM using a TGT1 tip calibration grating (NT-MDT Spectrum Instruments) featuring a grid of sharp tips.

The colloidal probe was then coated with GO in two steps. First, the probe sphere was functionalized with positive charges. Therefore, 1.98 g DI water was combined with 20 µl (3-trimethoxysilylpropyl) diethylenetriamine (Gelest). The probe was dipped in a droplet of this aqueous mixture for 30 minutes and then kept in the oven at 110 °C for 15 minutes. Second, the end of the probe was functionalized with GO (prepared using the improved Hummers method[49,50] and sonicated for 1 hour). Therefore, the probe was submerged in an aqueous GO dispersion for 30 minutes and then dried in a vacuum desiccator for one hour to remove water and stabilize the GO flakes. To reduce the GO coating, the probe was exposed to UV irradiation using a UV ozone cleaner (Novascan, PSD-UV) for 1 minute and then for 9 more minutes in $N_2$.

**Force spectroscopy.** Force curves were measured using an AFM (NT-MDT Spectrum Instruments, model: NTEGRA, AFM head model: SFC100LNTF) with a liquid cell (Model: MP3LC). First, the inverse optical lever sensitivity (IOLS) was determined by conducting force curves in water on the hard silicon substrate. Then, the force measurements on the heptane droplet were carried out with a typical range of 6 µm at a speed of 1 µm/s, measuring 2000 points per 1 µm.

The acquired raw data of the force curves were deflection (nA of diode photocurrent) vs. vertical piezo displacement (µm). The deflection was converted from nA to nm using the IOLS; to obtain forces in nN, the deflection values were multiplied by the spring constant $k$ of the cantilever ($k$=0.2 N/m). Some force curves showed a gradual decrease in the force up to a considerable separation distance due to the thermal drift of the cantilever. Therefore, baseline subtraction was performed for those curves, assuming that forces vanish at large separations.

## Results and Discussion

**Characterization of system.** Using the optical microscope in our AFM, we observed the pinned heptane droplets on the substrate in the water-filled liquid cell (Figure 2B); we found that they were relatively uniform in base diameter (≈20 µm). We measured the height of the droplet at points of a grid using force curves and determined the radius of curvature as ≈25 µm (see Supporting Information: S4). The immobilized heptane droplets were stable for a few hours. After closing the liquid cell of the AFM, the optical quality and contrast were reduced; however, we could still recognize the individual droplets. Simultaneously, we observed the AFM probe, which allowed us to choose a specific droplet to carry out force spectroscopy, as seen in Figure 2C. Specially, we selected droplets with approximately similar base diameters using the cantilever width as a reference.

An optical micrograph showing one of our homemade colloidal AFM probes is featured in Figure 3A; the 7 µm-diameter silica spheres ended up nearly perfectly centered at the end of the cantilever. Attaching spheres by this approach leads to a very well-defined, reproducible spherical probes that are the best geometry for theoretical descriptions. When we used coated cantilevers with reflective metal coatings, we occasionally observed the detachment of this metal layer; to avoid this issue and contamination of the liquid cell or sample, we worked exclusively with uncoated cantilevers. We characterized the surface structure of the attached colloidal particle using a calibration sample featuring sharp tips. The corresponding image in Figure 3B features the typical rough surface structure of such silica microparticles, with grains ~100 nm in diameter. This process was repeated after coating the probe with GO, and this image is shown in Figure 3C. The GO-coated probe topography displays the typical wrinkled morphology expected for GO, indicating successful attachment of the solvo-chemically produced graphene sheets.[50] SEM imaging of the probe also showed that the silica sphere is well wrapped by GO sheets while preserving its spherical shape (Figure 3D). Thickness of the GO sheets used for probe coating was characterized as ~1 nm (single layer) using AFM scanning and they were further imaged using TEM. (see Supporting Information: S1)

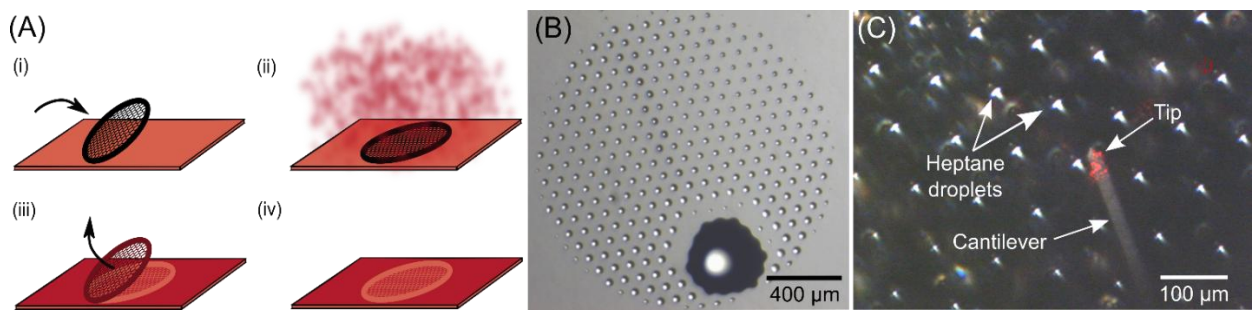

Figure 2. (A) Substrate preparation process. (i) A TEM grid is placed on a cleaned silicon substrate as a contact mask. (ii) A molecularly thin silane layer is deposited on the substrate except in areas covered by the TEM grid. (iii) TEM grid is removed. (iv) Substrate with hydrophobic patches in a hydrophilic background. (B) Optical image of immobilized heptane droplets in water (open AFM liquid cell). (C) Optical image of the droplets and cantilever during the experiment (closed AFM liquid cell). The heptane droplets can be seen brightly in the dark background. The colloidal tip sits under the end of the cantilever, and the laser beam is positioned on the end of the cantilever.

To tune the degree of hydrophobicity of each probe, we further used UV treatment to reduce the attached GO flakes. Each probe was first used without reduction ("GO") and then successively reduced for 1 minute ("rGO$_1$") and 10 minutes ("rGO$_{10}$") by exposing it to UV irradiation. The UV chamber was purged with $N_2$ to remove oxygen to avoid producing reactive radical oxygen atoms because the latter can cause further oxidation.[51,52] We confirmed that this method reduced GO by comparing the color of GO sheets before and after the treatment. As shown in Figures 3E and 3F, the treatment changed the material from light brown to dark brown, which is the expected color change when reducing GO to rGO. The images were acquired using identical illumination and imaging conditions. Correspondingly, they feature the same background color, and the observed color change represents their different reduction states. Reduction times longer than 10 minutes were not considered because excessive UV irradiation can introduce defects. The main advantage of using this approach was that it allowed us to carry out force spectroscopy with each probe at different levels of reduction and, thus, at different degrees of hydrophobicity. Importantly, the reduction does not alter the probe topography. Thus, we could be sure that the observed changes in the force curves were due to the hydrophobicity and not to changes in probe morphology.

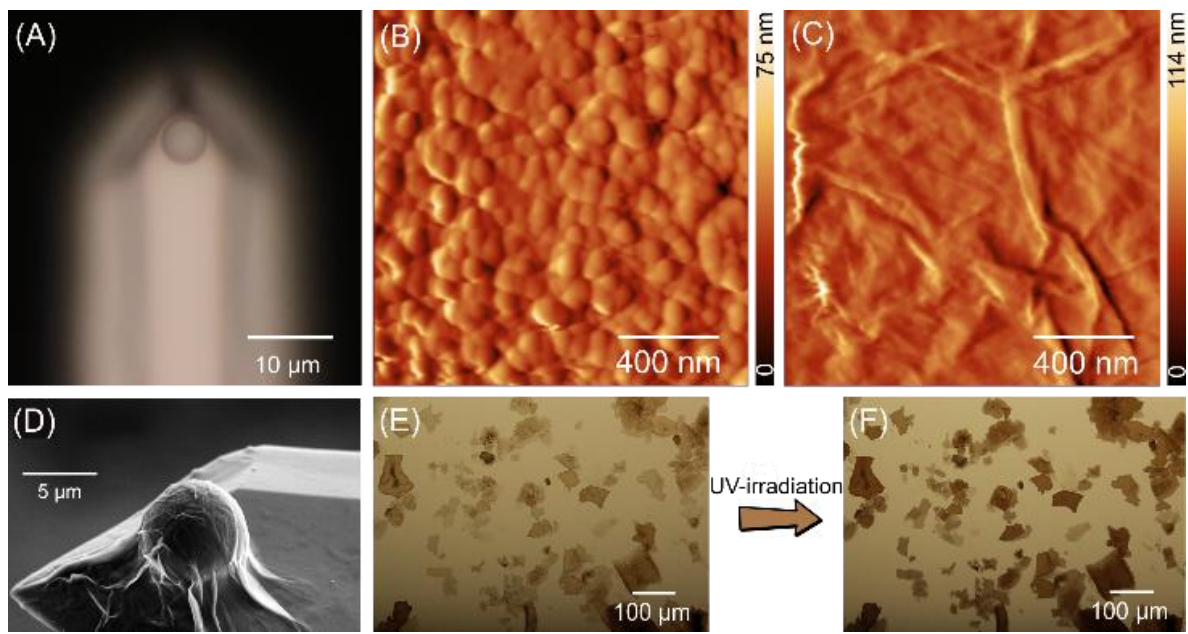

Figure 3. Characterization of custom-made AFM probes at different stages. (A) Optical image of a cantilever with a silica particle attached ("bare silica probe"). (B), (C) AFM images of the surfaces of bare silica probe (B) and GO-coated probe (C). (D) SEM image of GO coated probe (side view). (E), (F) Optical images of GO sheets on a glass slide (E) before and (F) after reduction.

**Force spectroscopy (FS) results and analysis.** A typical force–displacement curve of a GO-functionalized probe interacting with the water/heptane interface is shown in Figure 4. Schematic drawings 4*a*–4*g* illustrate how we picture the geometry of the interface at the different stages *a*–*g*, indicated by letters in the force curve. Magnified versions of some of the schematics are shown as 4*c'*–4*g'* in the bottom row of Figure 4. These show the contact angles clearly; the droplet and the sphere are drawn to scale. As expected, the force curve indicated no significant forces at large distances, meaning the heptane droplet remained in its original shape (*a*). As the piezo continued its approach, at ~0.92 μm piezo displacement (point *b*), the probe experienced a sudden "jump-in" event, typical in force spectroscopy when the slope of the interaction force becomes greater than the spring constant. After this event, the probe saw an attractive force of about −50 nN and established contact with the droplet (point *c*). Upon contact with the heptane droplet, a three-phase contact (TPC) line formed (see Figure 4*c'*). As the displacement of the piezo toward the sample continued, the area of the spherical probe covered by heptane increased, and the downward pointing forces were diminished. In between points *c* and *e*, the force curve featured several small jumps in the deflection, shown by the black arrows. These jumps indicated sudden motions of the TPC line across the spherical probe as the probe moved deeper into the heptane droplet, likely occurring when the TPC line moved across surface inhomogeneities of the probe, such as wrinkles in the GO coating. At point *d*, the force became zero, meaning neither the cantilever nor the water/heptane interface was deformed. Note that we took point *d* to define the zero point of our displacement. Continued reduction of the piezo displacement (motion of the probe toward the sample) then started causing a repulsive (positive) force (upward bending of the cantilever). This continued to point *e*, where the displacement minimum was reached, and the probe turned around to start the retracting phase. Correspondingly, the repulsive force also peaked at point *e*. Due to contact angle hysteresis, we do not expect the TPC line to move at first when the probe starts the return phase (moving from point *e* toward *g* in the force curve); instead, only the contact angle is expected to increase, as indicated in Figures 4*e'* and 4*f'*. This behavior is analogous to the movement of a droplet placed on an inclined surface when the incline angle is reversed. Reversal of the incline angle causes the (larger) advancing contact angle and the (smaller) receding contact angle to switch places. The receding back of the droplet then becomes the advancing front, but the contact line only starts moving once the advancing contact angle is reached. Thus, the first change in this reversal process is a corresponding deformation of the droplet, before the contact line starts moving. A similar change in contact angles is expected at the turnaround point in our force spectroscopy experiment. During the approach phase, the sphere is pushed deeper into the heptane and the water is pushed away; thus, we expect this to occur at the receding contact angle $\theta_r$ (Figure 4d') until the direction is reversed at point *e* (Figure 4e'). During the retract phase, the water-covered area of the sphere is ultimately growing again, which we expect to occur at the advancing contact angle $\theta_a$. However, since $\theta_a > \theta_r$, the first bit of the retracting sphere motion after the turnaround is translated into a deformation of the interface that increases the contact angle, until $\theta_a$ is reached (Figure 4f'), when the TPC line starts moving again, as suggested by Preuss et al.[29] At point *f*, the force became zero again, and the droplet was no longer deformed without a net external force. When the probe was further retracted, the TPC line started to move at point *f* in the force curve, indicated by a deviation of the force curve from the linear behavior indicated by the grey dashed line. At point *g*, the probe finally overcame the large adhesive force of −225 nN (more than four times the attractive force seen at point *c*) and released from the droplet at ~3.2 μm, where the force became zero again.

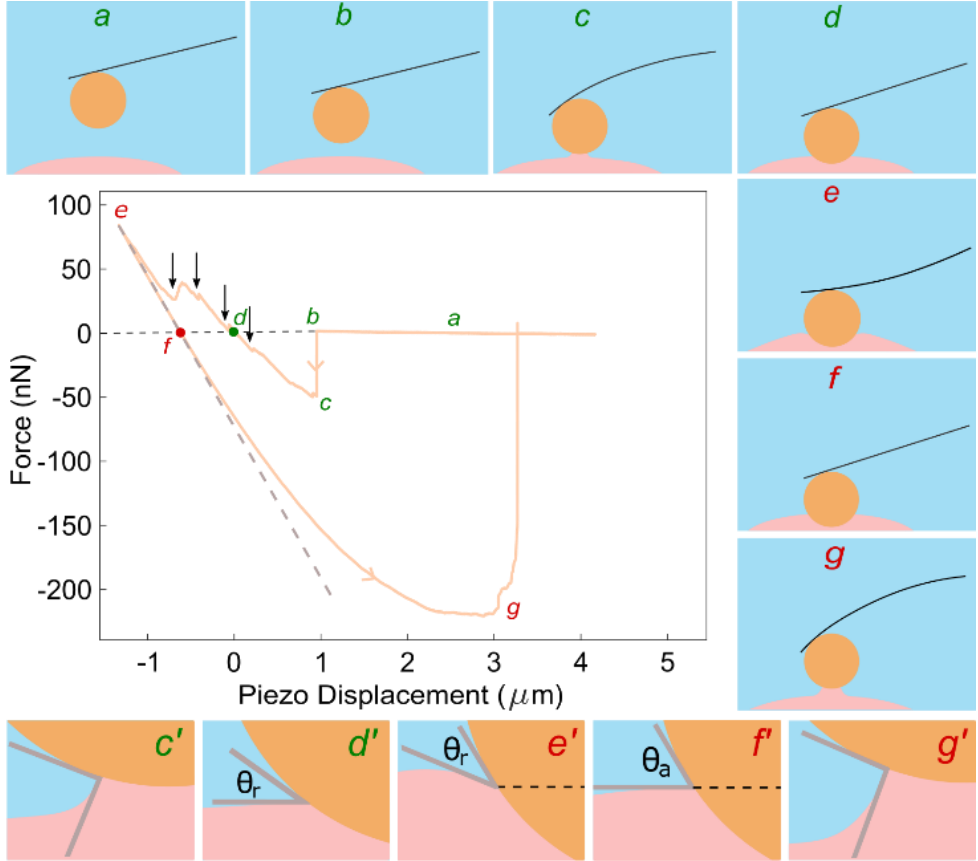

Figure 4: Force curve obtained between GO probe and heptane droplet in water. Black arrows indicate small jumps in the deflection. Green and red dots indicate points of zero force during approach and retract, respectively. Points *a–g* on the curve correspond to panels *a–g*. The gray dashed line (*e–f*) shows the data range used to calculate the system stiffness ($k_m$). Panels *a–g* demonstrate cantilever and droplet deformation during the force measurement. Panels *c'–g'* demonstrate the contact angle variation in water and correspond to panels *c–g*.

Subsequently, we carried out similar FS measurements using the very same probe after subjecting its GO coating to the UV reduction procedure for 1 minute and 10 minutes, respectively. Figure 5A shows representative force curves for GO (light orange), $rGO_1$ (orange), and $rGO_{10}$ (brown). These experiments were repeated with at least three different probes. The average values for the snap-on force (attractive force), snap-on displacement, and pull-off displacement were calculated from these curves, as shown in Figures 5B and C. As shown in Figure 5B, reducing GO for 1 minute to $rGO_1$ more than doubled the attractive force, on average, from 52 ± 9 nN to 124 ± 6 nN. Reduction by an additional 9 minutes increased the attractive force by a smaller amount (to 155 ± 35 nN). In accordance with the higher forces, the snap-on displacement systematically increased from GO through $rGO_1$ to $rGO_{10}$ in a similar way, from 1.0 ± 0.1 μm (GO) to 2 ± 0.4 μm ($rGO_1$) to 2.7 ± 0.1 μm ($rGO_{10}$), as shown in Figure 5C. Furthermore, for the re-duced GO, the probe had to be retracted by a greater displacement before it separated from the heptane droplet. The pull-off displacement was 3.8 ± 0.5 μm for GO, 5.8 ± 0.4 μm for rGO$_1$, and 6.1 ± 0.4 μm for rGO$_{10}$ (Figure 5C). All these three indicators, snap-on force, snap-on displacement, and pull-off displacement showed the same trend: the degree of interaction with the water/heptane interface significantly increased with the degree of GO reduction, with the biggest changes happening after only 1 minute of reduction.

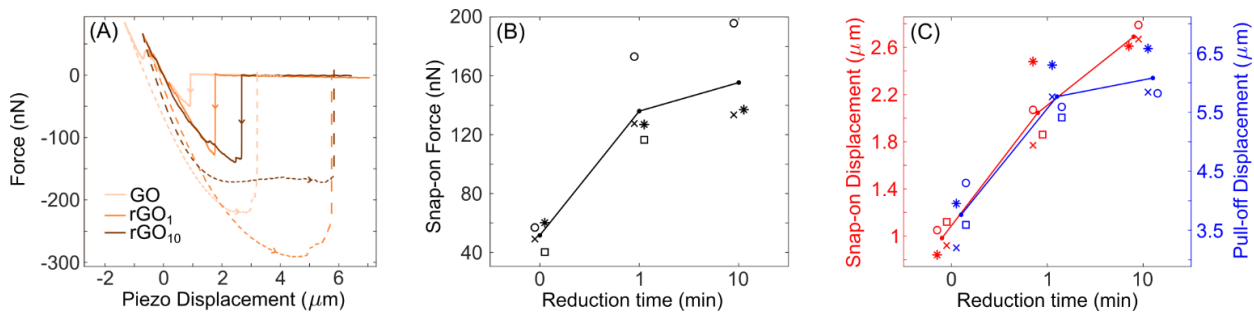

Figure 5: (A) Force spectroscopy results of an individual probe at different reduction levels (GO, rGO$_1$, rGO$_{10}$). The point where force becomes zero after snap-on was taken as the zero displacement for all force curves. Average snap-on force (B) and snap-on/pull-off displacements (C) for GO/GO$_1$/rGO$_{10}$. Different data point shapes indicate different probes. Solid dots connected by solid lines indicate averages.

Having established that the interactions of the probe with the water/oil interface depend on the reduction state of the GO coating, the next step was to test whether the hydrophobic nature of the surfaces indeed caused the observed attractive forces. The expected interactions in this scenario would be DLVO forces, which include electrostatic double-layer and van der Waals forces. According to a previous study, the water/heptane interface is negatively charged due to the spontaneous adsorption of hydroxyl ions.[53] Similarly, GO has a negative surface charge in water (isoelectric point: ~1.7),[54] primarily due to the dissociation of protons from the hydroxyl groups on the GO surface. Consequently, electrostatic forces between GO and the water/heptane interface should be repulsive; therefore, electrostatic forces can be excluded as the source of the observed attraction of GO to the water/heptane interface. A similar argument can be made for successively reduced probes rGO$_1$ and rGO$_{10}$, which would still be negatively charged, if to a lesser degree. This only leaves van der Waals and intrinsic or extrinsic hydrophobic forces as candidates to explain the observed attraction. Based on the positive value of the Hamaker constant between GO/graphene and heptane, van der Waals forces in our system would indeed be attractive and thus, a potential candidate for explaining our observations. We can distinguish van der Waals forces from other types of forces based on the interaction range, since van der Waals forces have a short range, typically on the order of a few nm.[19] Therefore, knowing the distance from which the AFM probe snaps onto the heptane droplet would be very helpful in distinguishing the different interactions.

Determining the exact snap-on distance is not trivial since both the droplet and the cantilever are deformed during this event and only the cantilever deformation $d$ is measured directly, whereas the droplet deformation needs to be inferred indirectly. To convert the measured force–displacement curves (Figured 5A) into force–distance curves,[55] where the distance represents the actual separation between the probe and the water/heptane interface, droplet deformation and cantilever deflection must be subtracted from the displacement for each data point.[29] To determine the unknown droplet deformation, we considered the droplet and the cantilever as a series of springs with effective stiffness $k_m$, which relates to the spring constants of the cantilever and the droplet, $k_c$, and $k_d$, as follows:

$$\frac{1}{k_m} = \frac{1}{k_d} + \frac{1}{k_c}. \tag{1}$$

Here, we consider the droplet as a Hookean spring, and its stiffness does not depend on the amount and the direction of the deformation. We determined the stiffness of the combined system using regions corresponding to points $e$–$f$ of the retract portion of the force curves, where the TPC line did not move, and thus, the total deformation of the combined spring must equal the negative change in piezo displacement. In other words, the slope of the force curves in this region equals $–k_m$. Based on the corresponding linear fit shown in Figure 4 as a dashed line and using the known cantilever stiffness $k_c$, we were able to calculate the droplet stiffness $k_d$ using Eq. (1). This allowed us to calculate the droplet deformation $s$ based on the measured force $F$ for each data point using $F=k_d·s$. Finally, we converted all the force–displacement curves in Figure 5A into force–distance curves shown in Figure 6A, effectively showing the probe's distance relative to the droplet's surface.

For the interpretation of the curves in Figure 6A, we further followed the previous work of Preuss et al., assuming that the colloidal probe first establishes contact with the interface in a single point.[29] Consequently, we considered this first contact point of the snap-on event (indicated by a vertical, dashed line in Figure 6A) as the zero distance. Positive distances on this calibrated force–distance curve indicate separations of the sphere from the droplet, ignoring any lo-cal deformations of the droplet, such as meniscus formation shown in Figures 4c', 4e', and 4g'. Similarly, negative distances represent the indentation depth of the probe across the liquid/liquid interface. Accordingly, we determined the separation between the probe and interface immediately before the snap-on event from this force–distance curve. For the probe shown in Figure 6A, the snap-in distance for GO was ~500 nm, and for $rGO_1$ and $rGO_{10}$ it was ~1 μm. The pull-off distances of $rGO_1$ and $rGO_{10}$ (5 μm and 4 μm, respectively) were higher than that of GO (~2.5 μm). We took force curves using at least three different probes for each experiment and applied the same conversion to force-distance curves in all cases. Then, we calculated the average snap-on and average pull-off distance values, as plotted in Figure 6B. The results aligned with the exemplary curves shown in Figure 6A: $rGO_1$ and $rGO_{10}$ consistently showed higher snap-on and pull-off distances than GO. This change confirms that rGO is attracted to the water/heptane interface more strongly than the GO upon approach; the same applies to the retraction, where a larger force is required to pull off the probe.

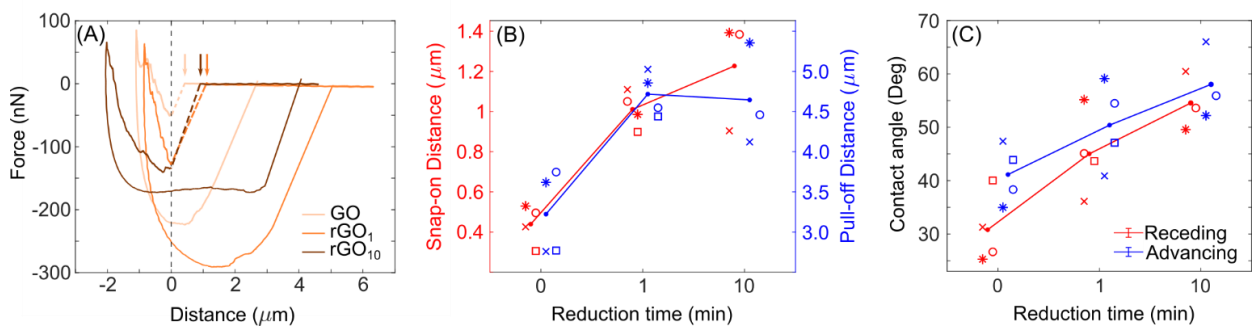

Figure 6: (A) Force–distance curves of GO, $rGO_1$, $rGO_{10}$. The black dashed line indicates the zero distance between the probe and the water/heptane interface, representing the first contact point during the snap-on event. Snap-on distances are indicated by arrows with corresponding colours. (B) Average snap-on distance and average pull-off distance. (C) Contact angle as a function of reduction time. The different data point shapes indicate different probes. Solid dots connected by solid lines indicate averages.

In all cases, the snap-on distances were significantly outside the typical van der Waals forces interaction range. This strongly suggests that GO and rGO are attracted to the water/heptane interface not due to van der Waals but to interactions caused by the hydrophobic nature of the surfaces. During the reduction of GO, polar functional groups are removed,[56] which makes the material increasingly hydrophobic. Thus, Figure 5B shows that the individual and average snap-on forces systematically increase as a function of hydrophobicity, which further supports our hypothesis that specifically the hydrophobic nature of the involved interfaces gives rise to the observed forces. Based on their interaction range, extrinsic effects of the surface hydrophobicity are expected to cause these forces.

From our force curves, we could further determine water contact angles on the probe surface in heptane. Therefore, we considered the zero-force points of our force curves, corresponding to points *d* and *f* in Figure 4, where no external force distorts the droplet and its contact angles. Figure 4 shows an offset in the distance variable of the two zero-force points between the approach and retract curves due to contact angle hysteresis. Consequently, we could determine receding and advancing contact angles $\theta_r$ and $\theta_a$ (see Figures 4d' and 4f') from the approach and retract curves, respectively. In general, the contact angle $\theta$ can be calculated by

$$\cos\theta = \frac{R-h}{R},\qquad(2)$$

where $R$ is the radius of the probe sphere and $h$ is the distance by which the TPC line moved across the sphere surface from the point of initial contact to the point of zero force (Figure 7A).[29] We were able to get $h$ from the force-distance curves by taking the absolute value of the distance since negative distances represent the amount of translation of the probe across the interface by our definition. The calculated contact angles using all the force curves were averaged and plotted in Figure 6C. As expected, the advancing water contact angle is always higher than the receding contact angle, regardless of the reduction level. Thus, for a given force, our probe is always deeper inside the heptane during the retraction than during the approach. Most importantly, both averages of the receding and advancing water contact angles increased with the reduction level, which corresponds with the increasing hydrophobicity of GO, in agreement with our observation that higher hydrophobicity (degree of GO reduction) increased both the snap-on forces and the snap-on distances and thus the interaction of the probe with the water/heptane interface.

After determining the water contact angles from the point of zero force, we now discuss how the force behaves as a function of indentation depth for the different probes. During the approach, the probe first snaps onto the water/heptane interface; as the approach is continued, it moves deeper across the interface, and accordingly, the TPC line moves across the sphere surface. The corresponding forces are represented by the approach curve for negative distances. In terms of forces, the repulsion increased as the probe was moved more deeply across the interface (Figure 7A). To further analyze this quantitatively, we considered the energy change when the probe moves more deeply across the interface. Equation 3 describes the surface energy change in the system as a function of $h$, the distance that the TPC line has moved from the initial contact point:

$$\Delta E = 2\pi R h \gamma_{gh} - 2\pi R h \gamma_{gw} - \pi(2Rh - h^2)\gamma_{wh}, \quad (3)$$

where the surface area of the sphere in contact with heptane as a function of $h$ is $2\pi Rh$. Thus, the first term describes the increase of sphere/heptane interfacial energy ($\gamma_{gh}$). At the same time, the sphere/water interfacial energy ($\gamma_{gw}$) decreases, as described by the second term. The third term represents the water/heptane interfacial area change due to the sphere movement. To obtain the corresponding force $F$, we differentiate Equation 3 with respect to $h$:

$$\frac{dE}{dh} = F = 2\pi R(\gamma_{gh} - \gamma_{gw} - \gamma_{wh}) + 2\pi \gamma_{wh} h. \quad (4)$$

Most importantly, this shows that the force changes linearly with respect to $h$. Interestingly, the proportionality constant ($2\pi\gamma_{wh}$) only depends on the interfacial energy between water and heptane ($\gamma_{wh}$) but not on the sphere size nor the material. The proportionality constant is positive, which means that the repulsion forces increase linearly as a function of indentation, which is entirely in line with our experimental observation: all the approach curves in Figure 7B representing different reduction levels of the probe feature a roughly linear region with similar slope, indicated by the blue, dashed line in Figure 7B. Interestingly, we directly determine the interfacial energy between water and heptane ($\gamma_{wh}$) from this experimentally measured slope, yielding $\gamma_{wh}$=40.4 mN/m, comparable to literature values.[57] Not only are our experiments fully in line with expectations based on our model, but our approach also represents a new experimental method to determine such interfacial energies, which only needs a microscopically small amount of the surface to be tested.

Finally, we discuss the impact of our work on understanding the origin of long-range interactions. Notably, we measured the range of these attractions between 0.5 μm and 1.2 μm (see Figure 6B). This aligns with the observations of Eriksson et al.,[3] who conducted experiments to measure interactions between a hydrophobic probe and a superhydrophobic surface using confocal microscopy. They observed the formation of an air bridge between the two surfaces at a distance of ~0.5 μm, which coincided with the onset of attractive forces. In their case, the air bridge formed between the hydrophobic probe and the air layer on the superhydrophobic surface. A significant difference between our system and theirs is that in our system there is only one solid sur-face (the colloidal AFM probe), which interacts with a liquid/liquid interface. It has been shown that air bubbles can form on a rough hydrophobic surface when this surface is immersed in water.[10,58] Therefore, we suggest that air bubbles also form on our hydrophobic (or amphiphilic) AFM probe coated with rGO or GO when immersed in the water. We also considered the feasibility of the formation of air bubbles or an air layer on the water/heptane interface; however, based on our estimates using surface energies, this is energetically not favorable: the sum of interfacial energies of the air/heptane interface (~20 mN/m)[59] and the water/air interface (~72 mN/m) is higher than the interfacial energy of the water/heptane interface (~50 mN/m).[57] Therefore, the bridging of air bubbles attached to both surfaces cannot be the origin of the observed long-range interactions. One possible way to explain our findings is to also consider the molecular rearrangement of water near hydrophobic surfaces, which has been hypothesized as the mechanism for the short-range, intrinsic hydrophobic interactions.[60,61] Donaldson et al.[13] have shown that the water molecules near the hydrophobic surfaces strongly fluctuate and are close to a liquid–vapor transition since they cannot form hydrogen bonds. So-called narrow depletion layers, ~0.1–0.6 nm thick, can thus be formed at hydrophobic solid/water, air/water, and oil/water interfaces.[62,63] Consequently, two such surfaces attract each other upon approach due to dewetting between the surfaces. However, this type of attraction would naturally be limited to a few nanometers, [13,61] which is far less than the range of attractive forces we observed.

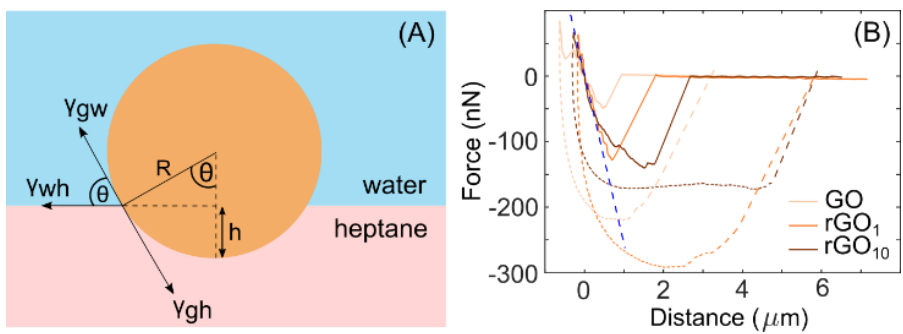

Figure 7: (A) The schematic diagram of the colloidal sphere at the water/heptane interface. The TPC line has moved from the initial contact point to $h$ distance (B) The approach curves at different reduction levels are overlaid to show the same force variation with respect to the distance.

Thus, the most likely possibility that explains our observation is bridging between existing air bubbles on the probe and a depletion layer at the water/heptane interface (Figure 8A). Since the air bubbles on the probe are also hydrophobic, they are likewise expected to be covered by a depletion layer (see Figure 8A). Thus, when the probe approaches the heptane droplet, the deple-tion layers on the bubble surface and water/heptane interface come into contact first (Figure 8B). This then leads to an attraction and the formation of an air bridge (Figure 8C). Finally, capillary forces further reduce the separation between the probe and the water/heptane surface, eventually leading to contact (Figure 8D). At this point, the cantilever is bent downward, and the heptane droplet is deformed by extending upward due to the attractive capillary force (Figure 8D). Furthermore, depending on the observed snap-on distances, we can estimate the height of the air bubbles to be in the range of 0.5–1.2 µm.

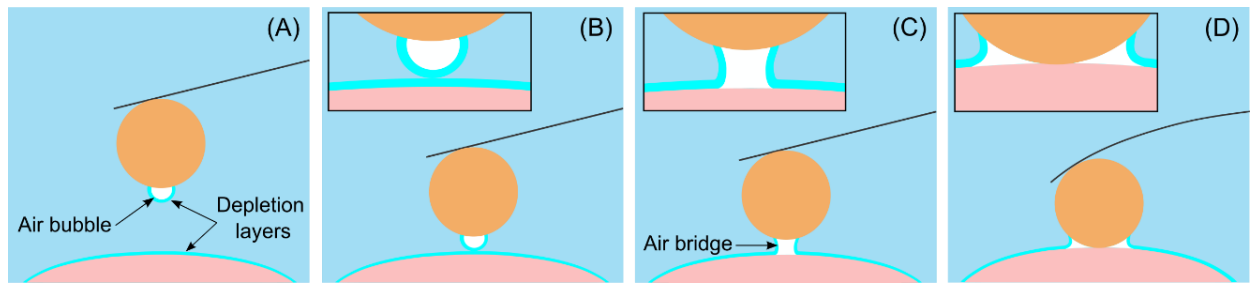

Figure 8: The mechanism of air bridging formation and the snap-on of the probe onto the droplet. (A) No interactions are observed when the probe–droplet separation is greater than the diameter of air bubbles on the probe. Depletion layers are present at the air/water interface and water/heptane interface. (B) When the probe approaches the heptane droplet, the depletion layers on the air/water interface and water/heptane interface come into contact. (C) An air bridge is formed between the probe and the heptane droplet, and a depletion layer is formed around the air bridge. (D) The probe and the droplet are pushed together just after forming the air bridge due to the capillary force. The attractive force bends the cantilever downward and deforms the heptane droplet by extending it upward. The insets of (B), (C), and (D) show in detail how the air bubble interacts with the heptane droplet.

We took many consecutive approach/retract cycles in which the probe had been fully retracted from the heptane bubble and found that the force curves showed exactly the same behavior. Based on our proposed mechanism, the air bubbles on the probe surface reform quickly after removing the probe from the heptane when the probe is again surrounded only by water. Figure 9A shows five force curves carried out consecutively at the same point with a duration of ~12 sec per curve. Accordingly, the probe was away from the heptane droplet for ~6 sec between retraction and approach.

The mechanism we proposed for the origin of interactions between a hydrophobic solid and a water/oil interface is equally applicable to explain the previous observations of non-DLVO, long-range interactions of hydrophobic solid particles with other hydrophobic particles or with water/gas interfaces. In other words, our model successfully explains the interactions of hydropho-bic particles with any hydrophobic body, and thus unifies and expands the realm of existing models.

Interestingly, in addition to the discussed hydrophobic attractive forces, we also observed a repulsive force just before the snap-on of the probe onto the droplet, featuring a decay length of ~30 nm. This repulsion started becoming detectable, i.e., the forces were greater than the noise at a distance of ~100 nm, which can be best seen in the magnified view of the data shown in Figure 9B. This made the repulsions clearly distinguishable from the periodic force oscillations due to optical interference (see Figure 9A inset and Supporting Information, S2). Similar repulsive forces were observed by Carambassis et al. between a hydrophobic colloidal probe and an air bubble in water.[64] To understand this repulsion between air bubbles and the heptane droplet, we considered (*i*) van der Waals (vdW) forces, (*ii*) electrostatic double layer (EDL) forces, or (*iii*) hydrodynamic drainage forces. The latter can arise due to a fluid's resistance when two surfaces approach each other quickly.[65,66] To assess whether vdW forces (*i*) are a likely explanation, we first calculated the Hamaker constant between an air bubble and a heptane droplet in water (see Supporting Information, S3). Interestingly, the obtained Hamaker constant is negative, indicating that the vdW forces are repulsive. However, when we calculated the actual force-vs-separation behavior based on our geometry and the Hamaker constant, we found that vdW are weaker and of shorter range than the observed repulsion (see Figure 9B and Supporting Information, S3 and S4). We thus ruled out vdW as a potential origin of the observed repulsion. To estimate the potential impact of EDL forces (*ii*), we estimated the Debye length of a hypothetic EDL repulsion between air bubbles and the heptane droplet in water for pH 5.5. We obtained ~240 nm, which also does not correlate with the decay length of the observed force (see Figure 9B). This leaves only hydrodynamic forces between the probe and the heptane droplet as a possible explanation for the observed forces. Carambassis et al. observed repulsive hydrodynamic forces featuring a decay length of about 9 nm.[64] This is within one order of magnitude of the observed repulsive forces, which featured a decay length of ~30 nm when an exponential fit was applied (see Figure 9B). The remaining discrepancy could be due to a difference in the specific geometry of our set-up relative to Carambassis et al.; these hydrodynamic forces critically depend on experimental conditions and are challenging to calculate.[67] Considering all these details, we think that hydrodynamic drainage forces are the most likely origin for the observed repulsive forces. Interestingly, not all force curves showed this repulsion before snap-on (Supporting Information, S2). The probes with the lowest snap-on distances at the same reduction level did not show repulsion. The lowest snap-on distances are shown when the contacting air bubble is smaller. This behavior agrees well with the nature of the hydrodynamic repulsion: force is low when the radius of curvature of the approaching body is low.[66]

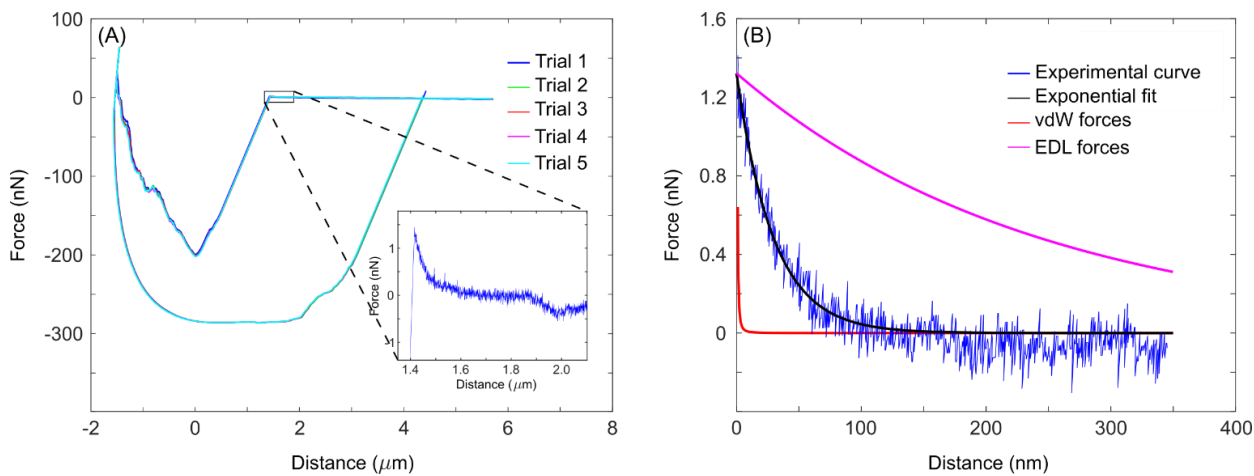

Figure 9 : (A) Five force curves obtained consecutively using a rGO$_{10}$ probe at the same spot on the heptane droplet are overlaid. They mostly show the same force behavior. The inset shows the zoomed-in area just before the snap-on of Trial 1 curve. (B) Comparison of experimental repulsive forces with modeled vdW and EDL forces. The black curve represents the exponential fit of the experimental curve. The vdW forces were modeled between an air bubble (Radius = 1.4 µm) and a heptane droplet (Radius = 25 µm) in water. The EDL forces were modeled using the calculated Debye length and the highest value of the exponential fit. The separation distance starts from zero as we consider the forces between an air bubble and the heptane surface.

Finally, our experimental results showed that the average range of the observed attraction increased with increasing hydrophobicity of the probe surface. According to the model of interaction we proposed, this would suggest that the height of the air bubbles on the probe also increased with the hydrophobicity level. Since the wettability of the probe surface decreased with increasing hydrophobicity, the formation of more and larger water bubbles would be expected for increasing hydrophobicity. Our observations are thus in line with our hypothesis of air bubble formation on the probe.

## Conclusions

In summary, we studied the interactions of hydrophobic solids with the liquid water/oil interface, which had not been investigated before. First, we performed force spectroscopy (FS) using AFM probes functionalized with GO as the solid surface. We then progressively increased the hydrophobicity of each probe by increasingly reducing the GO coating. This approach allowed us to prepare the same probe with different levels of hydrophobicity, thus eliminating tip shape effects, which otherwise notoriously limit the accuracy of AFM experiments.

Our FS results showed a significant attraction between the hydrophobic probes and the water/heptane interface, featuring large interaction ranges of 0.5–1.2 µm. Based on our analysis, the most likely mechanism for the origin of the observed long-range attractions is the bridging of air bubbles on the probe and the liquid–vapor depletion layer on the heptane droplet. The mechanism we propose is generally applicable to describe the non-DLVO, long-range interactions of hydrophobic particles with any hydrophobic body — in solid, liquid or gaseous form — and thus expands and unifies existing models of extrinsic hydrophobic interactions. The model is in line with our observation that the extrinsic hydrophobic interactions were stronger and featured longer ranges as the hydrophobicity of the probe increased. Our model further predicts increased interactions when a liquid with higher interfacial tension against water is used as the oil phase. Our experimental methods, on the other hand, provide a novel way of determining the interfacial energy at water/oil interfaces for microscopically small samples.

More specifically, our findings successfully explain the attraction between hydrophobic 2D materials and the water/oil interface, which have been successfully used to promote exfoliation of graphene and related materials. In these experiments, graphene and related materials are typically sonicated to produce single/few-layer sheets; air bubbles are likely formed on these 2D materials in this process, giving rise to their attraction to the water/oil interface. This phenomenon has subsequently been used to prepare graphene-stabilized emulsions, foams, and composite materials.[17,42] More broadly, we believe our insights will aid the understanding and enhancement of particle stabilized emulsion formation used in many applications.

## Associated Content

**Supporting Information.** More details regarding the following topics can be found at the end of this document, after the bibliography: GO flakes' thickness characterization, analysis of pre-contact repulsion, modeling van der Waals (vdW) forces, estimation of heptane droplet's radius of curvature.

**Data Availability.** The data that support the findings of this study are available at https://doi.org/10.7910/DVN/MEBRW1.[68]

## Author Contributions

Avishi Abeywickrama: experimental investigation, data analysis, methodology, visualization, writing – original draft preparation, review & editing. Douglas H. Adamson: writing – review & editing. Hannes C. Schniepp: conceptualization, supervision, funding acquisition, writing – review & editing.

## Funding Sources

This material is based upon work supported by the National Science Foundation under Award Nos. DMREF-1534428, DMREF-1535412, DMR-1905902, and DMR-2105158.

## Acknowledgements

The authors thank William & Mary Applied Research Center (ARC) for permitting access to their equipment.

## Abbreviations

GO, graphene oxide; DLVO, Derjaguin–Landau–Verwey–Overbeek; FS, Force spectroscopy; AFM, Atomic force microscopy; TEM, transmission electron microscopy; $rGO_1$, graphene oxide reduced for 1 minute; $rGO_{10}$, graphene oxide reduced for 10 minutes; TPC, three phase contact; EDL, electrostatic double layer; vdW, van der Waals.

## References

1. Meyer, E. E., Rosenberg, K. J. & Israelachvili, J. Recent progress in understanding hydrophobic interactions. *Proceedings of the National Academy of Sciences* **103**, 15739–15746 (2006).

2. Christenson, H. K. & Claesson, P. M. Cavitation and the Interaction Between Macroscopic Hydrophobic Surfaces. *Science* **239**, 390–392 (1988).

3. Eriksson, M. *et al.* Direct Observation of Gas Meniscus Formation on a Superhydrophobic Surface. *ACS Nano* (2019).

4. Attard, P. Long-range attraction between hydrophobic surfaces. *The Journal of Physical Chemistry* **93**, 6441–6444 (1989).

5. Miklavic, S. J., Chan, D. Y. C., White, L. R. & Healy, T. W. Double Layer Forces between Heterogeneous Charged Surfaces. *The Journal of Physical Chemistry* **98**, 9022–9032 (1994).

6. Podgornik, R. Electrostatic correlation forces between surfaces with surface specific ionic interactions. *The Journal of Chemical Physics* **91**, 5840–5849 (1989).

7. Tsao, Y. H., Evans, D. F. & Wennerstroem, H. Long-range attraction between a hydrophobic surface and a polar surface is stronger than that between two hydrophobic surfaces. *Langmuir* **9**, 779–785 (1993).

8. Nguyen, A. V., Nalaskowski, J., Miller, J. D. & Butt, H.-J. Attraction between hydrophobic surfaces studied by atomic force microscopy. *International Journal of Mineral Processing* **72**, 215–225 (2003).

9. Considine, R. F. & Drummond, C. J. Long-Range Force of Attraction between Solvophobic Surfaces in Water and Organic Liquids Containing Dissolved Air. *Langmuir* **16**, 631–635 (1999).

10. Hampton, M. A. & Nguyen, A. V. Nanobubbles and the nanobubble bridging capillary force. *Advances in Colloid and Interface Science* **154**, 30–55 (2010).

11. Ducker, W. A. & Mastropietro, D. Forces between extended hydrophobic solids: Is there a long-range hydrophobic force? *Current Opinion in Colloid &mathsemicolon Interface Science* **22**, 51–58 (2016).

12. Hato, M. Attractive Forces between Surfaces of Controlled "Hydrophobicity" across Water:? A Possible Range of "Hydrophobic Interactions" between Macroscopic Hydrophobic Surfaces across Water. *The Journal of Physical Chemistry* **100**, 1853018538 (1996).

13. Donaldson, S. H. *et al.* Developing a General Interaction Potential for Hydrophobic and Hydrophilic Interactions. *Langmuir* **31**, 2051–2064 (2014).

14. Xie, L. *et al.* Probing the Interaction between Air Bubble and Sphalerite Mineral Surface Using Atomic Force Microscope. *Langmuir* **31**, 2438–2446 (2015).

15. Xing, Y. *et al.* Recent experimental advances for understanding bubble-particle attachment in flotation. *Advances in Colloid and Interface Science* **246**, 105–132 (2017).

16. Xie, L., Shi, C., Cui, X. & Zeng, H. Surface Forces and Interaction Mechanisms of Emulsion Drops and Gas Bubbles in Complex Fluids. *Langmuir* **33**, 3911–3925 (2017).

17. Woltornist, S. J., Carrillo, J.-M. Y., Xu, T. O., Dobrynin, A. V. & Adamson, D. H. Polymer/Pristine Graphene Based Composites: From Emulsions to Strong, Electrically Conducting Foams. *Macromolecules* **48**, 687–693 (2015).

18. Mura, S., Nicolas, J. & Couvreur, P. Stimuli-responsive nanocarriers for drug delivery. *Nature Materials* **12**, 991–1003 (2013).

19. Israelachvili, J. N. *Intermolecular and Surface Forces*. (Elsevier LTD, Oxford: 2011).

20. Yang, H.-C., Hou, J., Chen, V. & Xu, Z.-K. Janus Membranes: Exploring Duality for Advanced Separation. *Angewandte Chemie International Edition* **55**, 13398–13407 (2016).

21. Wang, Z., Wang, Y. & Liu, G. Rapid and Efficient Separation of Oil from Oil-in-Water Emulsions Using a Janus Cotton Fabric. *Angewandte Chemie International Edition* **55**, 1291–1294 (2015).

22. Wang, S., Liu, K., Yao, X. & Jiang, L. Bioinspired Surfaces with Superwettability: New Insight on Theory, Design, and Applications. *Chemical Reviews* **115**, 8230–8293 (2015).

23. Yu, C., Zhu, X., Li, K., Cao, M. & Jiang, L. Manipulating Bubbles in Aqueous Environment via a Lubricant-Infused Slippery Surface. *Advanced Functional Materials* **27**, 1701605 (2017).

24. Rabinovich, Y. I. & Yoon, R.-H. Use of atomic force microscope for the measurements of hydrophobic forces. *Colloids and Surfaces A: Physicochemical and Engineering Aspects* **93**, 263–273 (1994).

25. Yoon, R.-H., Flinn, D. H. & Rabinovich, Y. I. Hydrophobic Interactions between Dissimilar Surfaces. *Journal of Colloid and Interface Science* **185**, 363–370 (1997).

26. Ishida, N., Kinoshita, N., Miyahara, M. & Higashitani, K. Effects of Hydrophobizing Methods of Surfaces on the Interaction in Aqueous Solutions. *Journal of Colloid and Interface Science* **216**, 387–393 (1999).

27. Ishida, N. Direct measurement of hydrophobic particle–bubble interactions in aqueous solutions by atomic force microscopy: Effect of particle hydrophobicity. *Colloids and Surfaces A: Physicochemical and Engineering Aspects* **300**, 293–299 (2007).

28. Shi, C., Chan, D. Y. C., Liu, Q. & Zeng, H. Probing the Hydrophobic Interaction between Air Bubbles and Partially Hydrophobic Surfaces Using Atomic Force Microscopy. *The Journal of Physical Chemistry C* **118**, 25000–25008 (2014).

29. Preuss, M. & Butt, H.-J. Direct Measurement of Particle-Bubble Interactions in Aqueous Electrolyte: Dependence on Surfactant. *Langmuir* **14**, 3164–3174 (1998).

30. Yang, Y. *et al.* An Overview of Pickering Emulsions: Solid-Particle Materials, Classification, Morphology, and Applications. *Frontiers in Pharmacology* **8**, (2017).

31. He, K., Li, Q., Li, Y., Li, B. & Liu, S. Water-insoluble dietary fibers from bamboo shoot used as plant food particles for the stabilization of O/W Pickering emulsion. *Food Chemistry* **310**, 125925 (2020).

32. Wu, F. *et al.* Investigation of the stability in Pickering emulsions preparation with commercial cosmetic ingredients. *Colloids and Surfaces A: Physicochemical and Engineering Aspects* **602**, 125082 (2020).

33. Lee, J. & Babadagli, T. Optimal design of pickering emulsions for heavy-oil recovery improvement. *Journal of Dispersion Science and Technology* **41**, 2048–2062 (2019).

34. Chevalier, Y. & Bolzinger, M.-A. Emulsions stabilized with solid nanoparticles: Pickering emulsions. *Colloids and Surfaces A: Physicochemical and Engineering Aspects* **439**, 23–34 (2013).

35. Pushkarova, R. A. & Horn, R. G. Bubble-Solid Interactions in Water and Electrolyte Solutions. *Langmuir* **24**, 8726–8734 (2008).

36. Helden, L., Dietrich, K. & Bechinger, C. Interactions of Colloidal Particles and Droplets with Water–Oil Interfaces Measured by Total Internal Reflection Microscopy. *Langmuir* **32**, 13752–13758 (2016).

37. Hartley, P. G., Grieser, F., Mulvaney, P. & Stevens, G. W. Surface Forces and Deformation at the Oil-Water Interface Probed Using AFM Force Measurement. *Langmuir* **15**, 7282–7289 (1999).

38. Ducker, W. A., Xu, Z. & Israelachvili, J. N. Measurements of Hydrophobic and DLVO Forces in Bubble-Surface Interactions in Aqueous Solutions. *Langmuir* **10**, 3279–3289 (1994).

39. Knebel, D., Sieber, M., Reichelt, R., Galla, H.-J. & Amrein, M. Scanning Force Microscopy at the Air-Water Interface of an Air Bubble Coated with Pulmonary Surfactant. *Biophysical Journal* **82**, 474–480 (2002).

40. Dagastine, R. R., Prieve, D. C. & White, L. R. Forces between a rigid probe particle and a liquid interface. *Journal of Colloid and Interface Science* **269**, 84–96 (2004).

41. Biswas, S. & Drzal, L. T. A Novel Approach to Create a Highly Ordered Monolayer Film of Graphene Nanosheets at the Liquid-Liquid Interface. *Nano Letters* **9**, 167–172 (2009).

42. Woltornist, S. J., Oyer, A. J., Carrillo, J.-M. Y., Dobrynin, A. V. & Adamson, D. H. Conductive Thin Films of Pristine Graphene by Solvent Interface Trapping. *ACS Nano* **7**, 7062–7066 (2013).

43. Kim, J. *et al.* Graphene Oxide Sheets at Interfaces. *Journal of the American Chemical Society* **132**, 8180–8186 (2010).

44. Kumar, H. V., Huang, K. Y.-S., Ward, S. P. & Adamson, D. H. Altering and investigating the surfactant properties of graphene oxide. *Journal of Colloid and Interface Science* **493**, 365–370 (2017).

45. Guo, P., Song, H. & Chen, X. Hollow graphene oxide spheres self-assembled by W/O emulsion. *Journal of Materials Chemistry* **20**, 4867 (2010).

46. Cote, L. J. *et al.* Graphene oxide as surfactant sheets. *Pure and Applied Chemistry* **83**, 95–110 (2010).

47. Gudarzi, M. M. & Sharif, F. Self assembly of graphene oxide at the liquid–liquid interface: A new route to the fabrication of graphene based composites. *Soft Matter* **7**, 3432 (2011).

48. Thickett, S. C. & Zetterlund, P. B. Graphene oxide (GO) nanosheets as oil-in-water emulsion stabilizers: Influence of oil phase polarity. *Journal of Colloid and Interface Science* **442**, 67–74 (2015).

49. Marcano, D. C. *et al.* Improved Synthesis of Graphene Oxide. *ACS Nano* **4**, 4806–4814 (2010).

50. Perera, D., Abeywickrama, A., Zen, F., Colavita, P. E. & Jayasundara, D. R. Evolution of oxygen functionalities in graphene oxide and its impact on structure and exfoliation: An oxidation time based study. *Materials Chemistry and Physics* **220**, 417–425 (2018).

51. Mulyana, Y., Uenuma, M., Ishikawa, Y. & Uraoka, Y. Reversible Oxidation of Graphene Through Ultraviolet/Ozone Treatment and Its Nonthermal Reduction through Ultraviolet Irradiation. *The Journal of Physical Chemistry C* **118**, 27372–27381 (2014).

52. Li, J., Qi, X., Hao, G., Ren, L. & Zhong, J. In-situ investigation of graphene oxide under UV irradiation: Evolution of work function. *AIP Advances* **5**, 067154 (2015).

53. Marinova, K. G. *et al.* Charging of Oil-Water Interfaces Due to Spontaneous Adsorption of Hydroxyl Ions. *Langmuir* **12**, 2045–2051 (1996).

54. Hu, X., Yu, Y., Wang, Y., Zhou, J. & Song, L. Separating nano graphene oxide from the residual strong-acid filtrate of the modified Hummers method with alkaline solution. *Applied Surface Science* **329**, 83–86 (2015).

55. Cappella, B. & Dietler, G. Force-distance curves by atomic force microscopy. *Surface Science Reports* **34**, 1–104 (1999).

56. Schniepp, H. C. *et al.* Functionalized Single Graphene Sheets Derived from Splitting Graphite Oxide. *The Journal of Physical Chemistry B* **110**, 8535–8539 (2006).

57. Zeppieri, S., Rodrıguez, J. & Ramos, A. L. L. de Interfacial Tension of Alkane mathplus Water Systems. *Journal of Chemical &mathsemicolon Engineering Data* **46**, 1086–1088 (2001).

58. Krasowska, M., Zawala, J. & Malysa, K. Air at hydrophobic surfaces and kinetics of three phase contact formation. *Advances in Colloid and Interface Science* **147-148**, 155–169 (2009).

59. Rolo, L. I., Caço, A. I., Queimada, A. J., Marrucho, I. M. & Coutinho, J. A. P. Surface Tension of Heptane, Decane, Hexadecane, Eicosane, and Some of Their Binary Mixtures. *Journal of Chemical &mathsemicolon Engineering Data* **47**, 1442–1445 (2002).

60. Chandler, D. Interfaces and the driving force of hydrophobic assembly. *Nature* **437**, 640–647 (2005).

61. Lum, K., Chandler, D. & Weeks, J. D. Hydrophobicity at Small and Large Length Scales. *The Journal of Physical Chemistry B* **103**, 4570–4577 (1999).

62. Bresme, F., Chacón, E., Tarazona, P. & Tay, K. Intrinsic Structure of Hydrophobic Surfaces: The Oil-Water Interface. *Physical Review Letters* **101**, (2008).

63. Bonn, M., Nagata, Y. & Backus, E. H. G. Molecular Structure and Dynamics of Water at the Water-Air Interface Studied with Surface-Specific Vibrational Spectroscopy. *Angewandte Chemie International Edition* **54**, 5560–5576 (2015).

64. Carambassis, A., Jonker, L. C., Attard, P. & Rutland, M. W. Forces Measured between Hydrophobic Surfaces due to a Submicroscopic Bridging Bubble. *Physical Review Letters* **80**, 5357–5360 (1998).

65. Gupta, R. & Fréchette, J. Measurement and Scaling of Hydrodynamic Interactions in the Presence of Draining Channels. *Langmuir* **28**, 14703–14712 (2012).

66. Kaveh, F., Ally, J., Kappl, M. & Butt, H.-J. Hydrodynamic Force between a Sphere and a Soft, Elastic Surface. *Langmuir* **30**, 11619–11624 (2014).

67. Manica, R. *et al.* Hydrodynamic forces involving deformable interfaces at nanometer separations. *Physics of Fluids* **20**, 032101 (2008).

68. Abeywickrama, A., Adamson, D. H. & Schniepp, H. C. Replication Data for: Long-Range Attraction between Graphene and Water/Oil Interfaces (2023). doi:10.7910/DVN/MEBRW1

# Supporting Information

## S1 GO flakes' thickness characterization

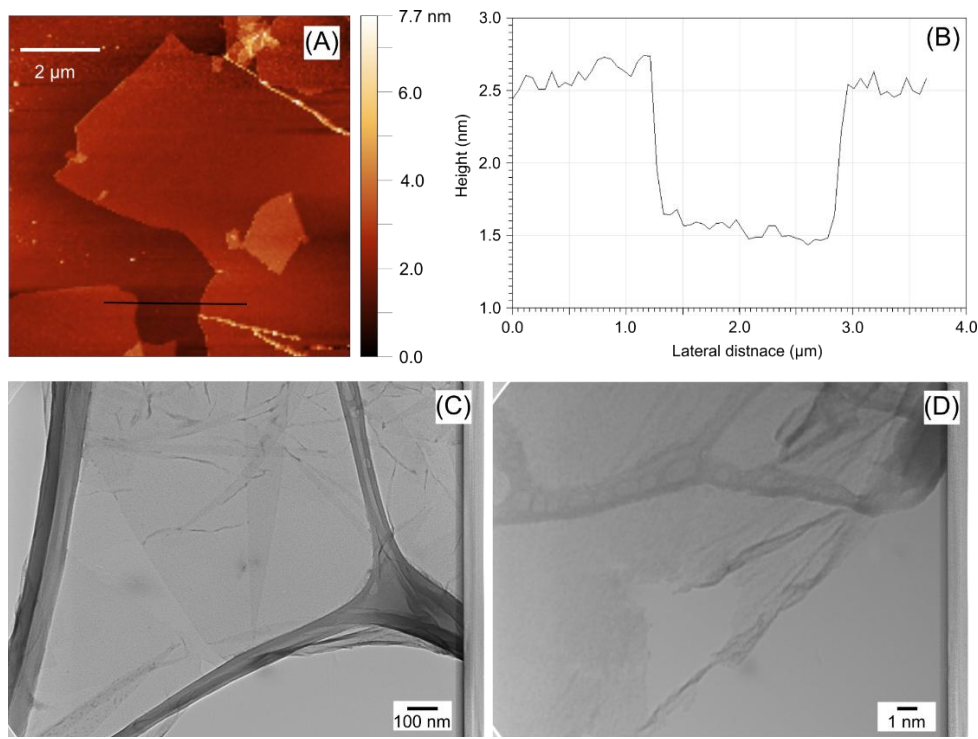

Figure S1: Thickness characterization of GO sheets used for probe functionalization. (A) Processed AFM image of GO sheets on mica substrate. (B) Height profile of the line drawn in (A). (C), (D) TEM images of GO sheets.

The GO dispersion was sonicated for one hour to exfoliate into few layer sheets. Before using them for functionalization we characterized the thickness of sonicated GO sheets using AFM (NT-MDT Spectrum Instruments, model: NTEGRA) and TEM (Philips FEI CM 10). AFM sample was prepared by spin coating (Laurell WS-400Bz-6NPP) a 2 µl droplet on mica substrate for 2 minutes at 2000 rpm. AFM scans were performed in semi-contact mode using an AFM probe with 40 N/m stiffness. Gwyddion software was used to flatten and analyze the AFM topography images (http://gwyddion.net/). The processed image is shown in Figure S1(A) and a line profile in Figure S1(B) confirms that the GO sheets are single layer with thickness of ~1 nm. The TEM samples were prepared by dipping a TEM grid (Ted Pella, INC., Support films, Lacey Formvar/Carbon, 300 mesh, Cu) in sonicated dispersion, and the obtained images at 80 kV and 245 kV are shown in Figure S1(C) and (D) respectively. These images also confirm that the GO sheets used for probe functionalization are single layer sheets.

## S2 Analysis of pre-contact repulsion

For some probes, we noticed a relatively small, but clearly observable repulsive force immediately before the snap-on. This becomes apparent when the corresponding region of the force curve is magnified, as shown in Figure S2(A) and S2(C). The periodic oscillations in the distance range 1.8–3.2 μm are due to well-understood optical inference effect and can be neglected. This repulsive force was not present for all probes, as shown in the example in Figure S2 (B) and (D); however, the frequency of observing this repulsion did not significantly depend on the degree of reduction of the probes.

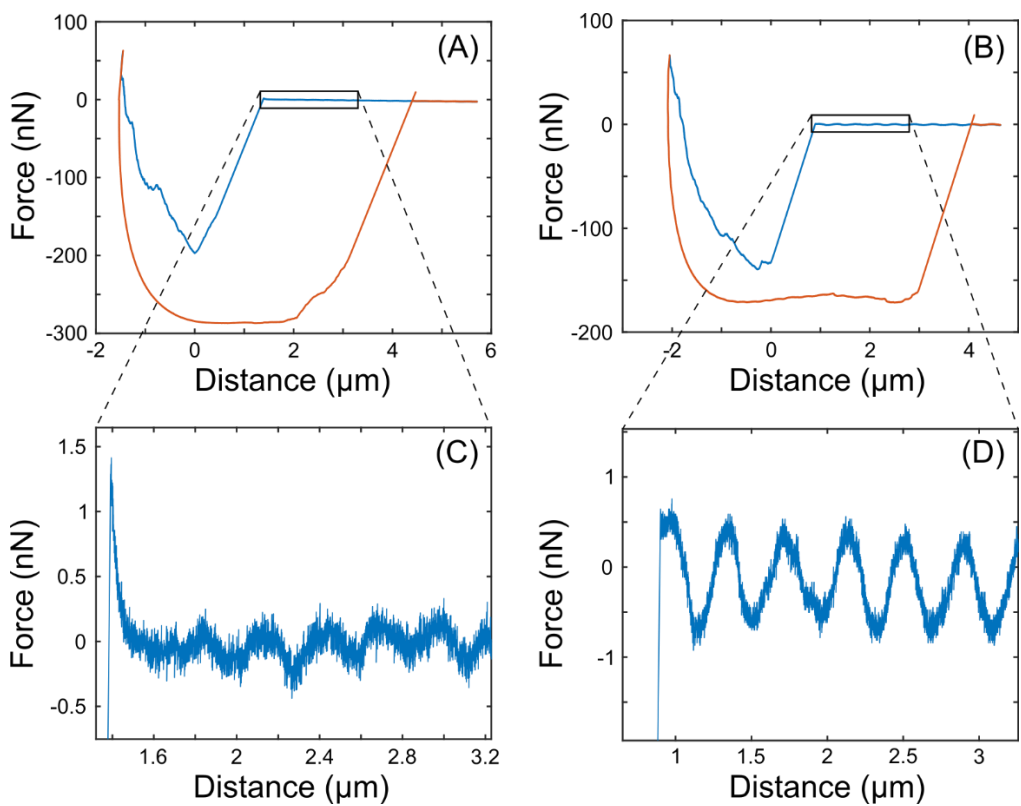

Figure S2: (A), (B) Force curves obtained using two probes at the same reduction level ($rGO_{10}$). (C) Magnified representation of the area indicated by the rectangle in panel (A), showing a repulsive force before the snap-on. The periodic force oscillations are due to optical interference. (D) Magnified representation of the area indicated by the rectangle in panel (B). The force is neutral before the snap-on event and only the force oscillations are present.

## S3 Modeling van der Waals (vdW) forces

To assess whether van der Waals forces, *F* between the air bubble and the heptane droplet may be the reason for the observed pre-snap-on repulsion, we first calculated the Hamaker constant, $A_H$ between air and heptane separated by water. Therefore, we used the following equation[19]

$$A_H = \frac{3}{4}kT\left(\frac{\varepsilon_1 - \varepsilon_3}{\varepsilon_1 + \varepsilon_3}\right)\left(\frac{\varepsilon_2 - \varepsilon_3}{\varepsilon_2 + \varepsilon_3}\right) + \frac{3h\nu_e}{8\sqrt{2}}\frac{(n_1^2 - n_3^2)(n_2^2 - n_3^2)}{(n_1^2 + n_3^2)^{1/2}(n_2^2 + n_3^2)^{1/2}\left\{(n_1^2 + n_3^2)^{1/2} + (n_2^2 + n_3^2)^{1/2}\right\}}$$

and then implemented it using the following MATLAB code:

```
k=1.38*10^-23; %Boltzman constant
T=298; %temperature (25C)
E1=1.9; %dielectric constant of heptane
E2=1; %dielectric constant of air
E3=78.4; %dielectric constant of water
n1=1.39; %refractive index of heptane
n2=1; %refractive index of air
n3=1.33; %refractive index of water
h=6.63*10^-34; %Planck's constant
v=3*10^15; %main electronic absorption frequency
A1=(3*k*T*(E1-E3)*(E2-E3))/(4*(E1+E3)*(E2+E3));
A2=3*h*v*(n1^2-n3^2)*(n2^2-n3^2)/(8*2^0.5);
A3=(n1^2+n3^2)^0.5*+(n2^2+n3^2)^0.5*((n1^2+n3^2)^0.5+(n2^2+n3^2)^0.5);
AAirHep=A1+(A2/A3) %Hamaker constant
```

Interestingly, we obtained a negative value, $A_H = -2.9 \times 10^{-21}$ J, which shows that the van der Waals force, *F* is indeed repulsive in this case and can thus modeled as[19]

$$F = \frac{-A_H}{6D^2}\left(\frac{R_1 R_2}{R_1 + R_2}\right),$$

where $R_1$, $R_2$ are the radii of curvature of the air bubble and the heptane droplet, $A_H$ is the Hamaker constant, and *D* is the separation distance between air bubble and the heptane droplet. The radius of curvature of the air bubble was assumed to be equivalent to its height (1.4 µm); the next section (S4) explains how we calculated the radius of curvature of the heptane droplet.

## S4 Estimation of heptane droplet's radius of curvature

Force measurements were carried out on a heptane droplet at different points following a grid pattern, featuring a separation of 3 µm between the grid points. Then, the snap-on displacement values were extracted from each force curve. These displacements were measured relative to the snap-on displacement next to the droplet, directly on the silicon substrate. In other words, the vertical distances of the snap-on positions to the silicon substrate were determined for different locations on the droplet surface, as shown in Figure S3. To model the curved droplet surface, a second-order polynomial fit was applied to the data points (Figure S3). The radius of curvature was determined from this fit as ≈25 µm.

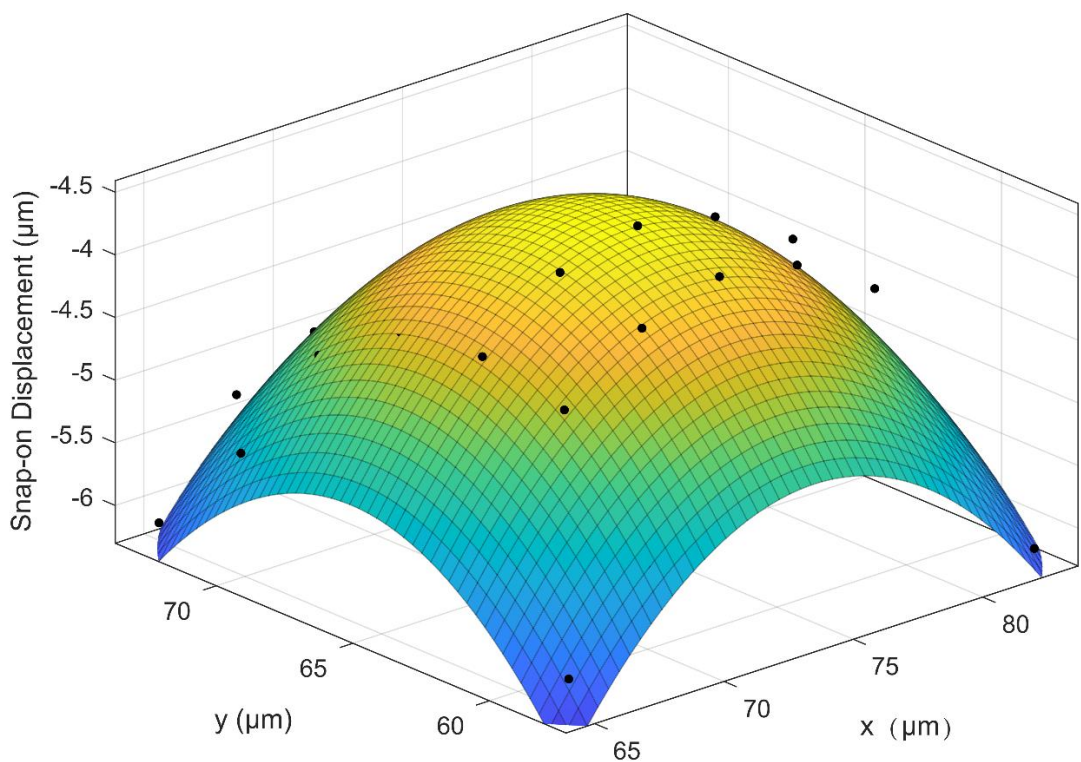

Figure S3: Modeled heptane droplet using snap-on displacement values at different points separated by 3 µm. Black dots indicate experimental values at corresponding points and some black dots are covered by the fit. The second order fit was applied in x and y directions using MATLAB.